\documentclass[12pt]{article} 
\pdfoutput=1
\usepackage{epsfig,graphicx,psfrag}
\usepackage{amsmath,amsfonts,stmaryrd,latexsym,amssymb}
\usepackage{url,hyperref,cite,relsize}
\usepackage{color}
\usepackage{slashed}
\usepackage{caption}
\usepackage{subcaption}

\numberwithin{equation}{section}

\newcommand{\bpm}{\begin{pmatrix}}
\newcommand{\epm}{\end{pmatrix}}

\def\met{E^{\rm miss}_{\rm T}}
\def\mt{M_{\rm T}}
\def\mttwo{M_{\rm T2}}
\def\mttw{M^W_{\rm T2}}

\def\chdi{{\boldmath $2\ell+\met~$}}
\def\chsemi{{\boldmath $1b1\ell+\met~$}}
\def\chbb{{\boldmath $2b+\met~$}}

\def\mst{m_{\tilde{t}}}
\def\msb{m_{\tilde{b}}}

\newcommand{\beq}{\begin{equation}}
\newcommand{\eeq}{\end{equation}}
\newcommand{\bea}{\begin{eqnarray}}
\newcommand{\eea}{\end{eqnarray}}

\begin{document}

\baselineskip=18pt \pagestyle{plain} \setcounter{page}{1}

\begin{flushright}
CALT-TH-2016-036 \\
DESY 16-101
\end{flushright}

\vspace*{1.2cm}

\begin{center}

{\Large \bf Exploring the nearly degenerate stop region with sbottom decays}
\\ [9mm]

{\normalsize \bf Haipeng An,$^{a,}$\footnote{~anhp@caltech.edu} ~  Jiayin Gu,$^{b,c,}$\footnote{~jiayin.gu@desy.de} ~  Lian-Tao Wang$\, ^{d,e,}$\footnote{~liantaow@uchicago.edu} \\ [4mm]
{\small {\it
$^a$ Walter Burke Institute for Theoretical Physics, \\ California Institute of Technology, Pasadena, CA 91125. \\ [2mm]
$^b$ Center for Future High Energy Physics, Institute of High Energy Physics, \\ Chinese Academy of Sciences, Beijing 100049, China. \\ [2mm]
$^c$ DESY, Notkestra{\ss}e 85, D-22607 Hamburg, Germany. \\[2mm]
$^d$ Enrico Fermi Institute, University of Chicago, Chicago, IL 60637.\\ [2mm]
$^e$ Kavli Institute for Cosmological Physics, University of Chicago, Chicago, IL 60637.
}}\\
}

\end{center}

\vspace*{0.2cm}

\begin{abstract}
A light stop with mass almost degenerate with the lightest neutralino  has important connections with both naturalness and dark matter relic abundance. This region is  also very hard to probe at colliders.  In this paper, we demonstrate the potential of searching for such stop particles at the LHC from sbottom decays, focusing on two channels with final states \chdi and \chsemi.   We found that, if the lightest sbottom has mass around or below 1\,TeV and has a significant branching ratio to decay to stop and $W$ ($\tilde{b} \to \tilde{t}\,W$), a stop almost degenerate with neutralino can be excluded up to about 500--600\,GeV at the 13\,TeV LHC with $300\,{\rm fb}^{-1}$ data.  The searches we propose are complementary to other SUSY searches at the LHC and could have the best sensitivity to the stop-bino coannihilation region.  Since they involve final states which have already been used in LHC searches, a reinterpretation of the search results already has sensitivity. Further optimization could deliver the full potential of these channels.
\end{abstract}

\newpage
{\small 
\tableofcontents}

\setcounter{footnote}{0}

\section{Introduction}
\label{sec:intro}

A light stop is essential for the naturalness of supersymmetry (SUSY). 
The stops have been extensively searched at the LHC. Traditional searches focus on the direct production of a stop pair followed by each stop decaying to the top quark and the lightest neutralino, $\tilde{t}\to t \, \chi$,\footnote{In this paper $\tilde{t}$ and $\tilde{b}$ always denote the lighter mass eigenstates, $\tilde{t}_1$ and $\tilde{b}_1$, unless specified otherwise.} while the lightest neutralino $\chi$ is the lightest superpartner (LSP).
The signal of these searches often includes large missing transverse momentum ($\met$) from the LSP.  
The current LHC bound for R-parity conserving SUSY models on stop mass is around $m_{\tilde{t}}\gtrsim 900\,$GeV, 
assuming $\tilde{t}\to t \, \chi$ and a sufficiently large mass gap between $m_{\tilde{t}}$ and $m_\chi$ \cite{CMS-PAS-SUS-16-028, CMS-PAS-SUS-16-029, CMS-PAS-SUS-16-030, ATLAS-CONF-2016-050, ATLAS-CONF-2016-077, Aad:2015pfx, Khachatryan:2016pup, Khachatryan:2016oia, CMS-PAS-SUS-16-002, Bradmiller-Feld:2016yvd, Aaboud:2016lwz}.\footnote{Stop may also decay to the lightest neutralino via an intermediate chargino or heavier neutralino, in which case the bounds on the stop mass is slightly weaker.}

The stops can still be significantly lighter than this bound if they are hiding in compressed regions $m_{\tilde{t}} \approx m_t + m_\chi$, $m_{\tilde t} \approx m_W + m_b+m_\chi$ and $m_{\tilde t}\approx m_\chi$, in which cases it is hard to discriminate the stop signal from standard model (SM) backgrounds or the products of the stop decay is too soft to be identified. Based on the stop-neutralino simplified model, searching strategies have been proposed to search for stops in these regions~\cite{Jezabek:1994qs,Brandenburg:2002xr,Han:2012fw,Aaltonen:2010nz,Abazov:2011qu,Abazov:2011ka,Abazov:2011mi,ATLAS:2012ao,Aad:2014pwa,Chatrchyan:2013wua,Aad:2014mfk,Choudhury:2012kn,Belanger:2013oka,Aad:2014nra, Khachatryan:2016pxa, Carena:2008mj,Bornhauser:2010mw,Ajaib:2011hs,Drees:2012dd,Dreiner:2012sh,Krizka:2012ah,Delgado:2012eu,Cohen:2013zla,Low:2014cba,Ferretti:2015dea,Khachatryan:2015wza,Hikasa:2015lma,Czakon:2014fka,Aad:2014kva,Rolbiecki:2015lsa,Dutta:2013gga,Drees:1993yr,Drees:1993uw,Martin:2008sv,Batell:2015zla,Grober:2014aha, Bai:2013ema, Cho:2014yma, Alves:2012ft,Hagiwara:2013tva,An:2015uwa,Macaluso:2015wja,Kobakhidze:2015scd,Cheng:2016mcw,Cheng:2016npb,Jackson:2016mfb,Kaufman:2015nda, Aaboud:2016tnv, Khachatryan:2016dvc, Goncalves:2016tft, Goncalves:2016nil}. \footnote{See also Ref.~\cite{Pierce:2016nwg} for a recent analysis on the implication of a light stop sector. } In R-parity conserving SUSY, the $m_{\tilde t} \approx m_\chi$ region is of special interests if the neutralino $\chi$ is mainly composed by the bino $\tilde B$. The reason is that the annihilation cross section of a pair of $\tilde B$ is small due to the lack of gauge interaction. Therefore, for $\tilde B$ to be a thermal dark matter candidate, a charged particle must be nearby to assist the annihilation. This region is thus called stop-bino coannihilation region~\cite{Ellis:2001nx}. This region is also not very well constrained by dark matter direct detection experiments~\cite{Berlin:2015njh}.  According to the numerical simulation with micrOmegas4.2~\cite{Belanger:2014vza}, for sub-TeV bino-like dark matter, a mass difference $m_{\tilde t_1} - m_{\chi} \approx 30\,$GeV is required to obtain the measured relic abundance.  In the simulations of this work, we fix this mass gap to be 30 GeV.  The sensitivities of collider search discussed in this paper are slightly better if the mass gap is smaller.  In this compressed region, the stop has two main decay channels, one is the flavor-conserving four-body decay through off-shell top quark and W boson ($\tilde{t}\to bW^*\chi \to bl\nu \chi / bjj \chi$) and the other is the flavor-changing two-body decay to a charm quark ($\tilde{t}\to c\,\chi$). The decay rate of the flavor-changing channel depends strongly on the flavor structure of the squark sector, whereas the rate of the four-body channel depends only on the mixing angle between the left and right handed stop. It turns out that with $m_{\tilde t_1} - m_{\chi} = 30$\,GeV, the four-body channel alone makes the stop decay promptly~\cite{Grober:2014aha}.

\begin{figure}
\centering
\includegraphics[height=5cm]{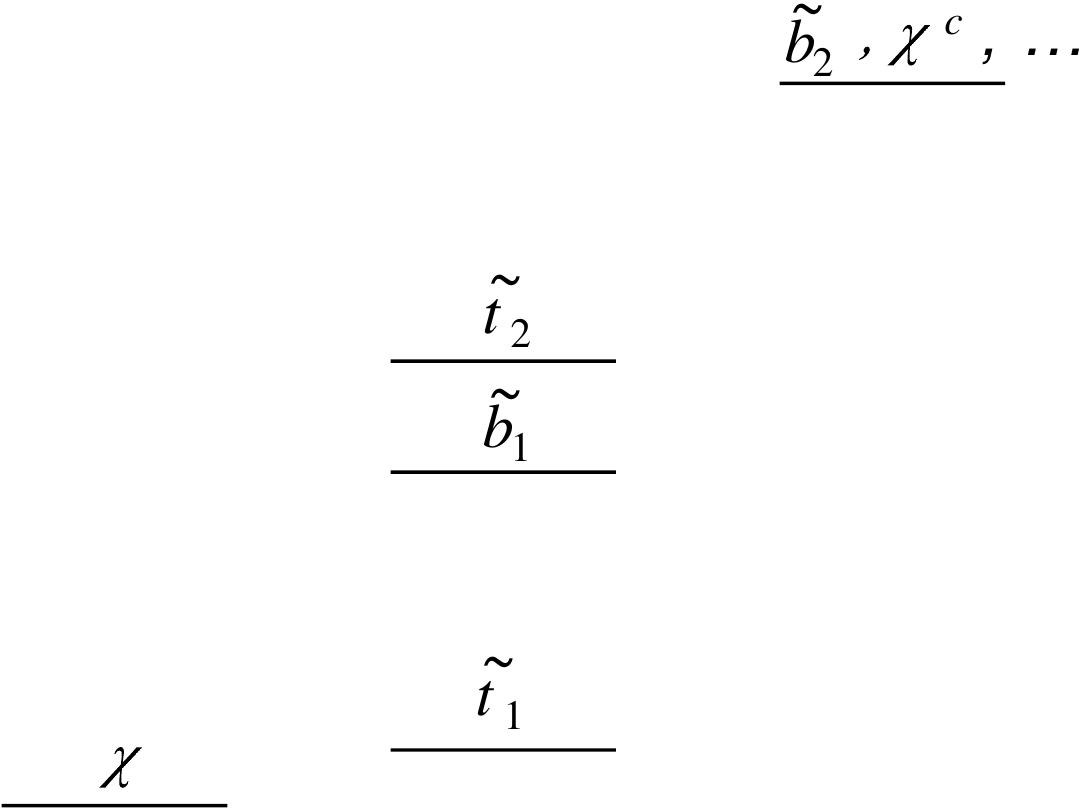}
\caption{Spectrum of SUSY partners of the stop-bino coannihilation region.}
\label{fig:spectrum}
\end{figure}

In this paper, we draw attention to a couple of additional useful search channels using sbottom decays, to further probe this nearly degenerate region. Naturalness prefers the second stop not to be too heavy.
Due to the doublet nature of the left handed quarks, the masses of the left handed sbottom is connected to the mass of the left handed stop. The mixing between the left and right handed stops usually makes the mass of the second stop heavier than the left handed sbottom. To minimize the flavor violation induced by the squark sector, the mixing between the left and right handed sbottoms is usually assumed to be suppressed by the mass of the bottom quark. Therefore, we can decouple the right handed sbottom in this work.  Our search strategy relies on a significant mass gap between the lightest stop and sbottom, which we obtain by assuming $m_{\tilde{t}_R}$ is sufficiently smaller than $m_{\tilde{t}_L} (= m_{\tilde{b}_L})$, and the lightest stop is mostly right-handed.~\footnote{ A large mass gap could also be generated by a very large stop $A$ term even if $m_{\tilde{t}_R}\approx m_{\tilde{t}_L}$, but with such a large $A$ term also comes the risk of spontaneously breaking $SU(3)_{\rm c}$. In this scenario, the decay $\tilde{b} \to \tilde{t}\,W$ would dominate which makes our case even stronger. } To simplify the study we also assume the winos and the Higgsinos are decoupled and the lightest neutralino is pure bino.  The spectrum of the SUSY particles is shown in Fig.~\ref{fig:spectrum}. In this simplified scenario, the lighter sbottom $\tilde b_1$ has two decay channels
\begin{eqnarray}
\tilde b_1 &\rightarrow& W + \tilde t_1 \ ,\nonumber \\
\tilde b_1 &\rightarrow& b + \chi \ ,
\end{eqnarray}
with decay rates
\begin{eqnarray}\label{eq:gammas}
\Gamma_1 &\equiv& \Gamma_{\tilde b_1 \rightarrow W \, \tilde t_1} = \frac{g_2^2 \sin^2\theta_t \cos^2\theta_b}{32\pi} \frac{[(m_{\tilde b}^2 - (m_{\tilde t}+m_W)^2)(m_{\tilde b}^2 - (m_{\tilde t}-m_W)^2)]^{3/2}}{ m_W^2 m_{\tilde b}^3} \ , \nonumber \\
\Gamma_2 &\equiv& \Gamma_{\tilde b_1 \rightarrow b \, \chi} ~ = \frac{g_1^2}{32\pi} \frac{(m_{\tilde b}^2 - m_\chi^2)^2}{m_{\tilde b}^3} 4 \left[ \left(-\frac{1}{3}\right)^2 \sin^2\theta_b + \left(\frac{1}{6}\right)^2 \cos^2\theta_b\right] \ ,
\end{eqnarray}
where in calculating $\Gamma_2$ we neglect the mass of the bottom quark. The stop and sbottom mixing angles are defined as
\begin{equation}
\bpm \tilde{t}_1 \\ \tilde{t}_2 \epm = \bpm \cos\theta_t & \sin\theta_t \\ -\sin\theta_t & \cos\theta_t \epm \bpm \tilde{t}_R \\ \tilde{t}_L \epm   \,, \hspace{1cm}
\bpm \tilde{b}_1 \\ \tilde{b}_2 \epm = \bpm \cos\theta_b & \sin\theta_b \\ -\sin\theta_b & \cos\theta_b \epm \bpm \tilde{b}_L \\ \tilde{b}_R \epm   \,.\label{eq:massmix}
\end{equation}
In the limit $m_{\tilde b}^2-m_{\tilde t}^2  \gg m_W^2$, $\Gamma_1$ is seemingly enhanced by the factor $m_{\tilde b}^2/m_W^2$ due to the longitudinal contribution. However, the stop mixing angle vanishes if the electroweak symmetry is unbroken. Therefore, the stop mixing angle $\theta_t$ is secretly proportional to $m_W$. In the limit $ A_t v \ll m_{\tilde t_2}^2 \approx m_{\tilde b}^2$, we have
\beq
\sin\theta_t \approx \frac{\sqrt{2} A_t m_W}{g_2 m_{\tilde b}^2} \ ,
\eeq
where $A_t$ is the $A$ term for the stops. Assuming $A_b$ is suppressed by $m_b$ for the sake of flavor physics constraints, we have $\cos\theta_b\approx 1$. Therefore, $\Gamma_1$ and $\Gamma_2$ in Eq.~(\ref{eq:gammas}) can be simplified as
\beq\label{eq:gammas2}
\Gamma_1 \approx \frac{A_t^2}{16\pi m_{\tilde b}} \ , \;\;\;\; \Gamma_2 \approx \frac{\alpha_{\rm em}m_{\tilde b}}{72 \cos^2\theta_W} \ .
\eeq
The proportionality of $\Gamma_1$ to $A_t^2$ can also be inferred from the goldstone equivalence theorem. The traditional sbottom search based on the sbottom-neutralino simplified model assumes the sbottom decays 100\% to $b$ and the neutralino. However, as from Eq.~(\ref{eq:gammas2}) if $A_t$ is comparable to $m_{\tilde b}$, $\Gamma_1/\Gamma_2$ can be as large as ${\cal O}(100)$. This region is also favored by the Higgs mass. On the other hand, in some specific SUSY breaking models (e.g. gauge mediation models) $A_t$ is one-loop order suppressed compared to other soft SUSY breaking parameters. In this case, $\Gamma_1 \ll \Gamma_2$. Therefore, when searching for the signal from sbottoms, it is important to consider both of the two decay channels. 

We will focus on studying the potential of the sbottom decay channels shown in {\bf (a)}  and {\bf (b)} of Fig.~\ref{fig:signal}. We apply relatively straightforward cuts to demonstrate that these channels can lead to interesting reach with an integrated luminosity of 300 fb$^{-1}$ at the 13\,TeV LHC. A more careful optimization of the kinematical selection  and more realistic simulation are needed to determine the ultimate reach. This is beyond the scope of this paper. We also present the reach in the more ``conventional" sbottom search channel in shown in {\bf (c)} of Fig.~\ref{fig:signal} to illustrate the complementarity between these channels. We would like to emphasize that even in the parameter region in which  {\bf (c)} has a better reach, the new channels  {\bf (a)}  and {\bf (b)}  studied in this paper is still useful in the case of a discovery since they directly probe the presence of the stop.

The rest of this paper is organized as follows:  In Section~\ref{sec:channel}, we discuss the main search channels and the corresponding backgrounds.   
In Section~\ref{sec:study}, we state the selection cuts for each channel and show the results of a few case studies.  In Section~\ref{sec:reach}, we show the exclusion regions in the parameter space of the 13\,TeV LHC with $300\,{\rm fb}^{-1}$ data and compare the reaches of different channels.  The conclusion is drawn in Section~\ref{sec:conclusion}.


\section{Search channels and backgrounds}
\label{sec:channel}

\begin{figure}[t]
\centering
    \begin{subfigure}[b]{0.3\textwidth}
        \includegraphics[trim = 6cm 20cm 7cm 2cm, clip, height=5cm]{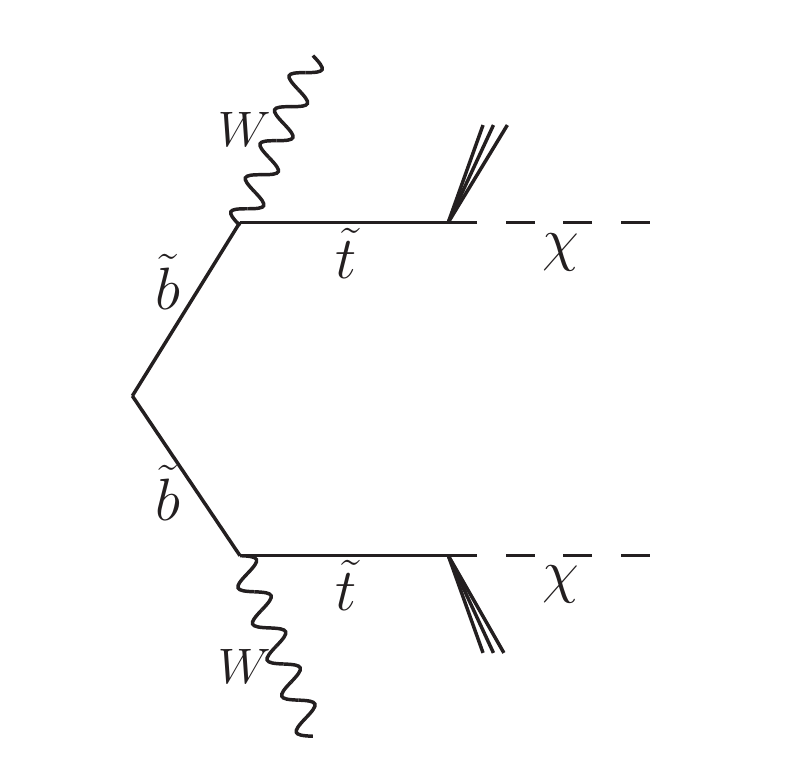}
        \caption{}
        \label{fig:siga}
    \end{subfigure}~
    \begin{subfigure}[b]{0.3\textwidth}
        \includegraphics[trim = 6cm 20cm 7cm 2cm, clip, height=5cm]{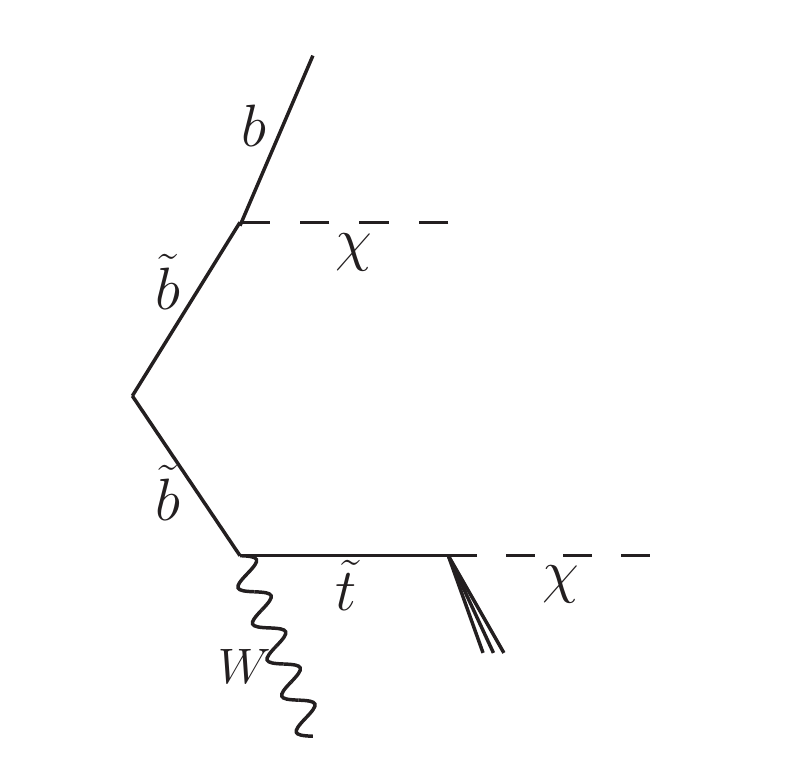}
        \caption{}
        \label{fig:sigb}
    \end{subfigure}~
    \begin{subfigure}[b]{0.3\textwidth}
        \includegraphics[trim = 6cm 20cm 9cm 2cm, clip, height=5cm]{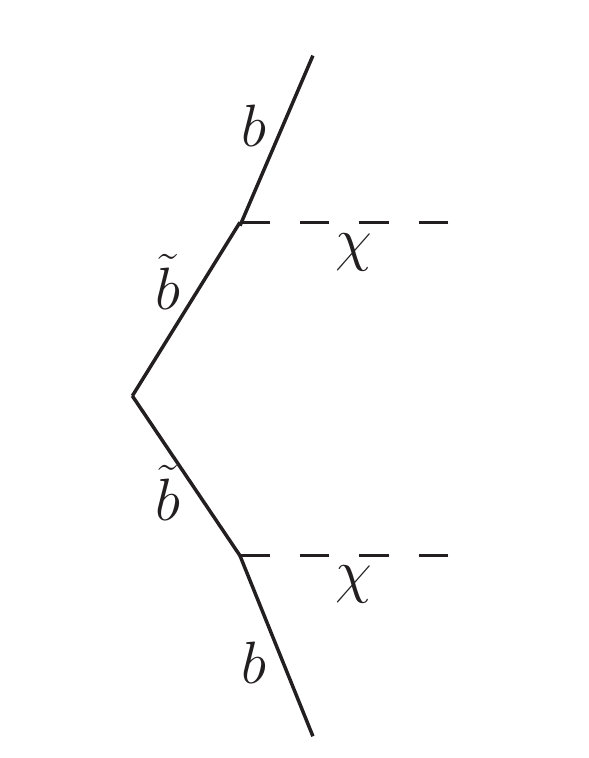}
        \caption{}
        \label{fig:sigc}
    \end{subfigure}
\caption{Three ways for a sbottom pair to decay for the scenario studied in this paper:  {\bf (a)} symmetric decay of $\tilde{b} \to \tilde{t}\,W$, {\bf (b)} asymmetric decay, and {\bf (c)} symmetric decay of $\tilde{b}\to b\,\chi$. }
\label{fig:signal}
\end{figure}

With two decay channels $\tilde{b} \to \tilde{t}\,W$ and $\tilde{b}\to b\,\chi$, a pair of sbottoms produced at the LHC has three ways to decay, as shown in Fig.~\ref{fig:signal}.  
The symmetric decay chain of $\tilde{b}\to b\,\chi$ in Fig.~\ref{fig:sigc} has already been searched at the LHC in the channel with final states  \chbb under the assumption of $100\%$ branching ratio (BR), and sbottom with mass up to 800\,GeV are excluded for $m_\chi \lesssim 360\,$GeV \cite{Aaboud:2016nwl}.  With a smaller branching ratio, the reach of this channel is significantly weaker.  Here our main interest is in the decay chains that involves the stop, namely, the symmetric decay chain in Fig.~\ref{fig:siga} and the asymmetric decay chain in Fig.~\ref{fig:sigb}.  As such, we will focus on two channels, one with final states of two opposite sign leptons and $\met$ ({\boldmath $2\ell+\met$}), and the other with one hard b-jet, one lepton and $\met$ ({\boldmath $1b1\ell+\met$}).  These two channels are studied in details in Section~\ref{sec:study}, while in Section~\ref{sec:reach} we compare their reaches together with the one of the \chbb channel for different sbottom branching ratios.

The \chdi channel is designed to pick up the the symmetric decay chain in Fig.~\ref{fig:siga} with both $W$s decaying leptonically, and should be the optimal search channel if the decay $\tilde{b} \to \tilde{t}\,W$ dominates.
This channel has been searched at the LHC for the searches of sleptons and electroweakinos \cite{Aad:2014vma, Khachatryan:2014qwa}, for which the main background is top quark pair production ($t\bar{t}$\,)  
and diboson ($WW/WZ/ZZ$).  We also found the $t\bar{t}Z$ events to have a significant contribution to the SM backgrounds after imposing our selection cuts.

The \chsemi channel is designed for the asymmetric decay chain in Fig.~\ref{fig:sigb}, but could also pick up some events from the symmetric decay chain in Fig.~\ref{fig:siga} with one $W$ decaying hadronically, if the event happens to have a hard b-jet.        
This channel is similar to the direct search of stop pair in the semileptonic channel \cite{Aad:2015pfx, Khachatryan:2016pup, Aaboud:2016lwz}, where the main backgrounds include $t\bar{t}$, $tW$, $W+\,$jets, diboson and $ttZ$.  We expect this channel to be useful if the branching ratios of $\tilde{b} \to \tilde{t}\,W$ and $\tilde{b}\to b\,\chi$ are comparable.

In principle, one could also search in the channel with final states of one lepton, $\met$ and one or two hard jets with no b-jets ({\boldmath $1\ell+\rm{jets}+\met$}), which could come from either Fig.~\ref{fig:siga} with one $W$ decaying hadronically or Fig.~\ref{fig:sigb} if the b-jet is not tagged.  While this channel could contain significant amount of signal events, the backgrounds are also large  and more complicated. In this paper, we focus on the simpler leptonic channels as an initial assessment of the potential of these new decay channels.


\section{Simulation procedure and event selection}
\label{sec:study}

For both signal and backgrounds, the events are generated at parton level using Madgraph5 \cite{Alwall:2014hca}, followed by parton showering with PYTHIA6.4 \cite{Sjostrand:2006za}.  The detector simulation is performed with Delphes \cite{deFavereau:2013fsa} in which the b-tagging efficiency is from Ref.~\cite{ATL-PHYS-PUB-2015-022}.  In particular, the b-tagging efficiency is within 60\%--70\% for $p_T$ in the range 50--300\,GeV, while the mis-tag rate is below 15\% for a charm jet and below 0.5\% for a light jet with $p_T<400$\,GeV. We use the above procedure to generate the events of sbottom pair production and then rescale the cross section to the values from the NLO+NLL calculation in Ref.~\cite{Borschensky:2014cia, sbcs}. \footnote{ Jet matching is not performed for signal events as scanning the parameter space with jet matching requires a huge amount of computing power.  Since for signal events the hard b-jets and charged leptons are from the decay of the sbottoms, we do not expect the behavior of ISR jets to have a strong effect on the behavior of the signal.  We have also checked with specific signal benchmark points that this is indeed the case.  } For $t\bar t$, single top and $W,Z$+jets events, the MLM matching procedure is also employed. For $t\bar t$ events, the total cross section is scaled to the NNLO+NNLL result given in Ref.~\cite{Czakon:2011xx, ttcs}.  For single top (including $tW$) events, the total cross section is scaled to the NLO results in Ref.~\cite{singletopcs}.  For diboson events, the total cross section is scaled to the NLO result in Ref.~\cite{Campbell:2011bn}.  For $t\bar{t}Z$ events, we scale the cross section to the central value of the recent measurement in Ref.~\cite{Aaboud:2016xve}.

We present the details of our collider study in this section, including the selection cuts for each channel and the results of a few case studies.  The signal samples listed Table~\ref{tab:s1234} are used for the case studies.  Signal S1\,\&\,S2 has ${\rm BR}(\tilde{b}\to \tilde{t}\,W)=0.9$ and are ideal for the \chdi channel, while S3\,\&\,S4 has ${\rm BR}(\tilde{b}\to \tilde{t}\,W)=0.5$ which is better covered by the \chsemi channel.  For  S1\,\&\,S3, we assume the stop only decays to charm and neutralino, $\tilde{t}\to c\, \chi$; for S2\,\&\,S4, we assume that the stop only goes through the 4-body decay, $\tilde{t}\to bW^*\chi \to bl\nu \chi / bjj \chi$.  
The mass spectra in Table~\ref{tab:s1234} are chosen to roughly correspond to the ``best reach" of the two channels, which are shown later in Section~\ref{sec:reach}.

\begin{table}
\centering
\begin{tabular}{|c||c|c|c|c|c|c|} \hline
signal   & $\sigma {\rm [pb]}$ & $m_{\tilde{b}} {\rm [GeV]}$ & $m_{\tilde{t}} {\rm [GeV]}$ & $m_\chi {\rm [GeV]}$ & BR($\tilde{b}\to \tilde{t}\,W$) & $\tilde{t}$ decay   \\  \hline\hline
S1 & 0.00615  & 1000  &  600  &  570  &  0.9  &  $c\,\chi$   \\  \hline
S2 & 0.00615  & 1000  &  600 &   570  &  0.9  &   $ bl\nu \chi / bjj \chi$  \\  \hline
S3 & 0.0129    &  900   &  500 &   470  &  0.5  &   $c\,\chi$  \\  \hline
S4 & 0.0129    &  900   &  500 &   470  &  0.5  &   $ bl\nu \chi / bjj \chi$  \\  \hline
\end{tabular}
\caption{Signal samples used for the case studies in the \chdi channel (S1\,\&\,S2) and the  \chsemi  channel (S3\,\&\,S4).}
\label{tab:s1234}
\end{table}

\subsection{ {\boldmath $2\ell+\met$} channel }

{\bf Selection cuts:} For an event to pass the cut, we require it to have $\met>150\,$GeV and contain exactly 2 leptons with opposite charge.  We require the scalar sum of the $p_T$ of the two leptons to be larger than $200\,$GeV.  We also apply a b-veto by requiring the event to have no b-jet with $p_T>50\,$GeV.  The requirement on $p_T$ of b-jets could prevent one from removing signal events with soft b-jets from stop decays.  We require the invariant mass of the lepton pair ($m_{ll}$) to be larger than $20\,$GeV to remove potential backgrounds from low mass resonances.  If the two leptons have the same flavor, we further require their invariant mass to be at least 20\,GeV away from the $Z$ boson mass.   
A stringent cut around the $Z$ resonance helps remove the $ZZ$ background with $ZZ \to \ell^+ \ell^- \nu \bar{\nu}$, which cannot be efficiently removed by the $\mttwo$ variable \cite{Lester:1999tx, Barr:2003rg} due to the different event topology.  In order to remove events with a large $\met$ coming from mis-measurements of jet energy, we require that the azumithal angle between the missing transverse momentum and any jet with $p_T>50\,$GeV to satisfy $|\phi_{\rm MET}-\phi_j|>0.2$.  Finally, we require the $\mttwo$ of the lepton pair to be larger than $150\,$GeV.

\begin{figure}
\centering
\includegraphics[height=6cm]{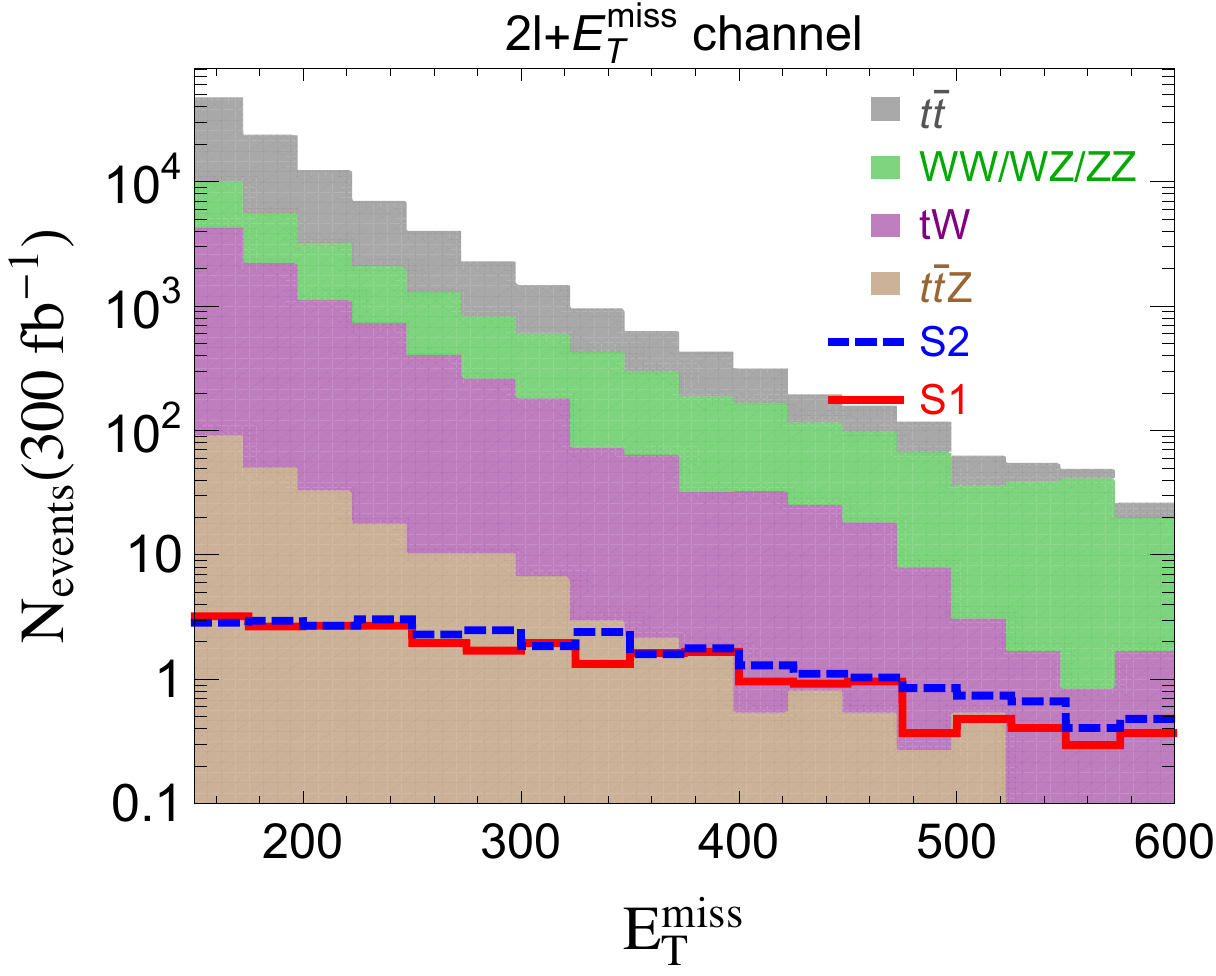} \hspace{0.5cm}
\includegraphics[height=6cm]{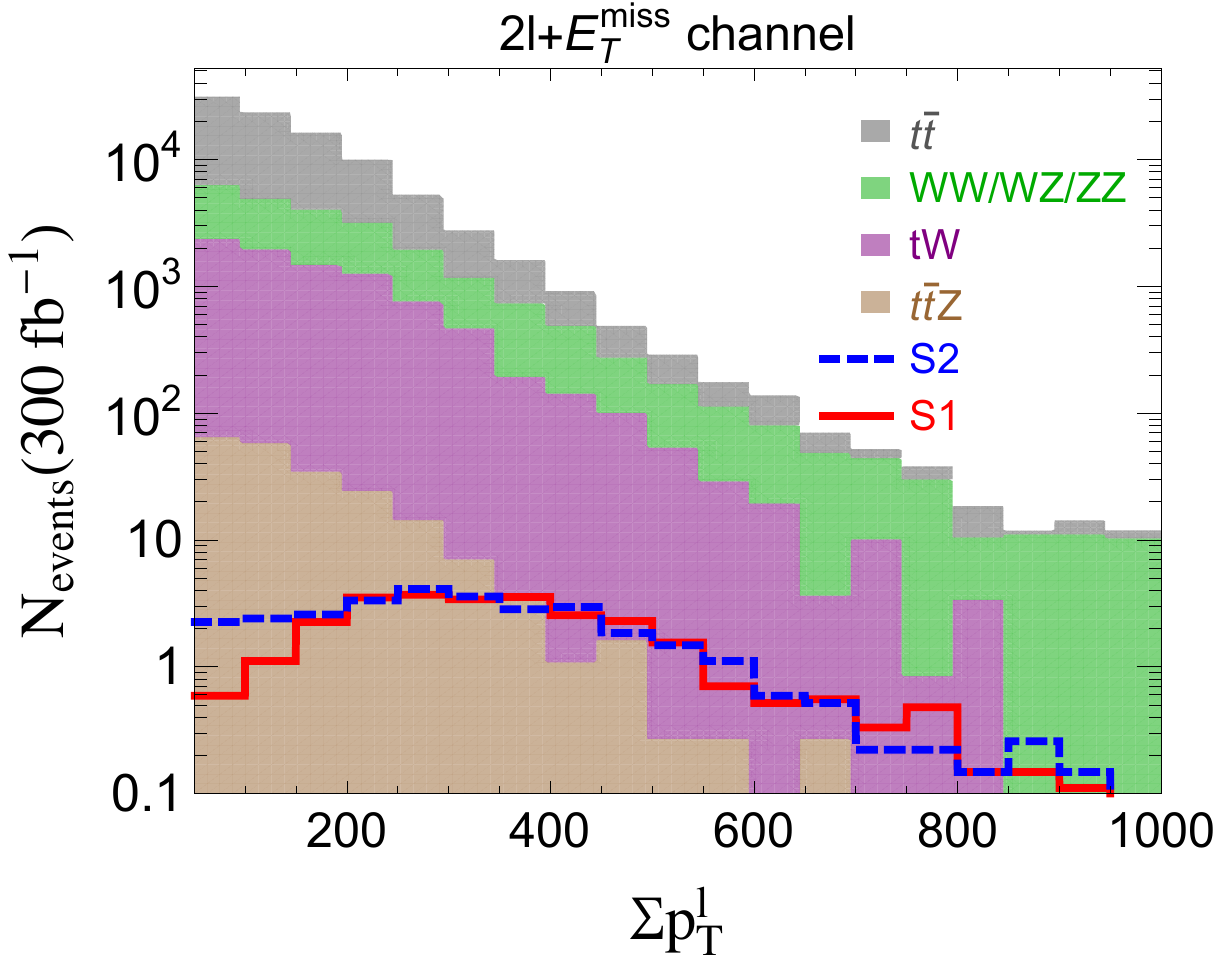}  \\ \vspace{0.1cm}
\includegraphics[height=6cm]{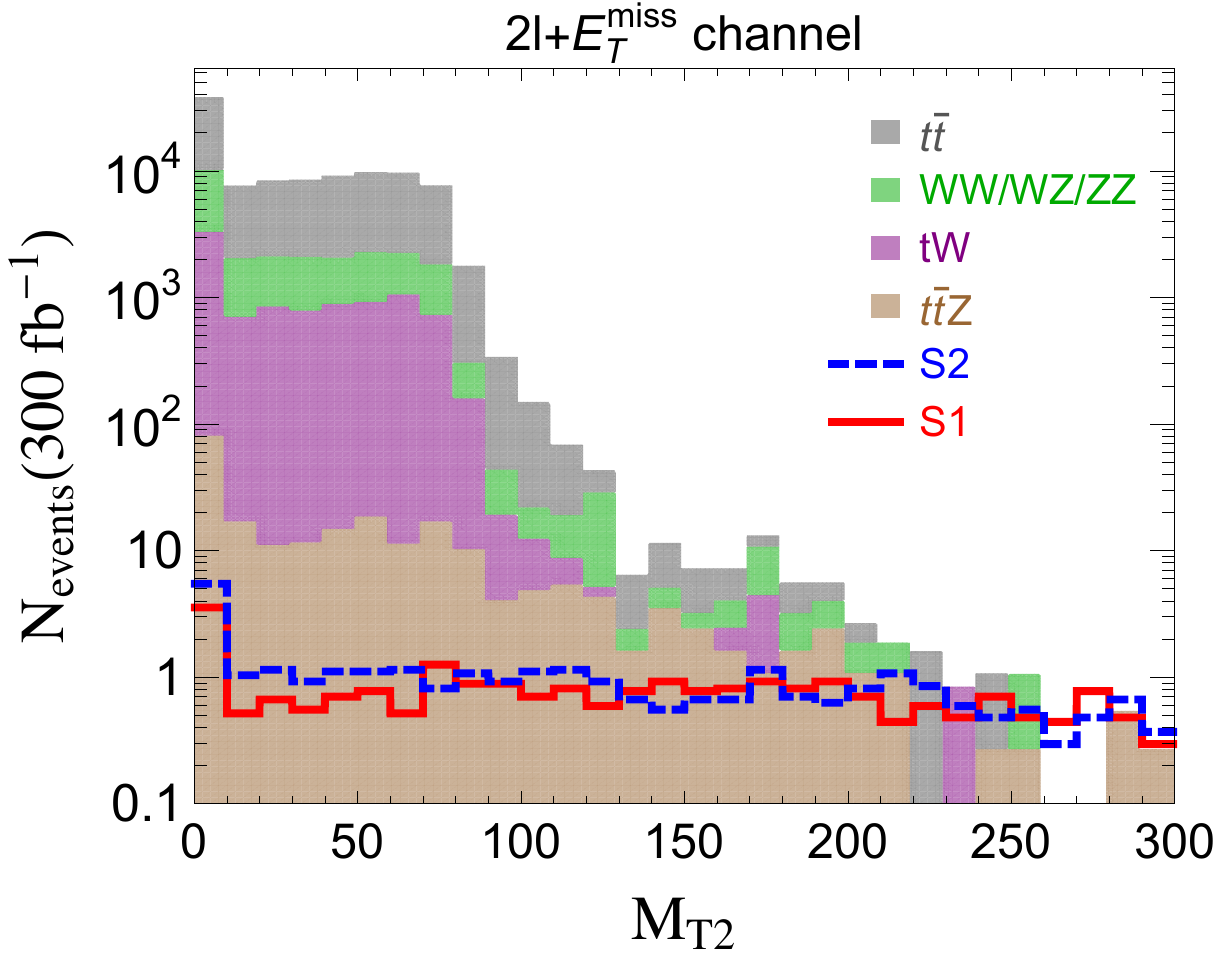} 
\caption{Distributions of $\met$ ({\bf top left}), $\sum p^l_T$ ({\bf top right}) and $\mttwo$ ({\bf bottom}) of the \chdi channel for signal sample S1 and the major backgrounds.  $\sum p^l_T$ is the scalar sum of the $p_T$ of the two leptons.  To illustrate the usefulness of the variables, the cuts \{$\sum p^l_T>200\,$GeV , $\mttwo>150\,$GeV\} are removed. The number of events correspond to $300\,{\rm fb}^{-1}$ at the 13\,TeV LHC.}
\label{fig:dilep}
\end{figure}

The distributions of $\met$, $\sum p^l_T$ (scalar sum of the $p_T$ of the two leptons) and $\mttwo$ are shown in Fig.~\ref{fig:dilep} for signal S1, S2 and the major backgrounds.  In Fig.~\ref{fig:dilep} one could clearly see the endpoint feature of the $\mttwo$ distribution of the $t\bar{t}$, $WW/WZ/ZZ$ and $tW$ backgrounds.~\footnote{Note that most $ZZ$ background are removed by the lepton invariant mass cut.  If this cut is not imposed, a significant amount of $ZZ$ background will have $\mttwo\gtrsim m_W$. }  A cut on $\mttwo$ with a value much larger than the $W$ mass is very efficient at removing these backgrounds.  On the other hand, the $t\bar{t}Z$ background has additional neutrinos and does not have the endpoint feature.  While it has a much smaller cross section, after the $\mttwo$ cut we found it to be comparable with other major backgrounds.  The numbers of signals and backgrounds after the selection cuts and the corresponding $s/\sqrt{b}$ for $300\,{\rm fb}^{-1}$ data are shown in Table~\ref{tab:sig2l}.  
Comparing the results of S1 and S2, one could see that the decay channel of the light stop has a rather small impact on the reach, due to the high jet and lepton threshold we choose to use.

\begin{table}
\centering
\begin{tabular}{c|cc} 
  & \# of events ($300\,{\rm fb}^{-1}$) & $s/\sqrt{b}$  \\ \hline 
S1 &  13  &  2.0   \\
S2 &  12  & 2.0   \\ \hline\hline
$t\bar{t}$ & 11  &  \\
$WW/WZ/ZZ$  &  15  &  \\
 $t\bar{t}Z$   &  8.4  &   \\
 $tW$  & 5.1   &    \\ \hline
 total SM  &  40  &  
\end{tabular}
\caption{ Number of events of signal and backgrounds and the corresponding $s/\sqrt{b}$ after all of the selection cuts for the \chdi channel with $300\,{\rm fb}^{-1}$ data.  The details of signal samples S1\,\&\,S2 are listed in Table~\ref{tab:s1234}.  All the generated backgrounds are included in the row ``total SM."      
}
\label{tab:sig2l}
\end{table}
%

\subsection{{\boldmath $1b1\ell+\met$} channel}

{\bf Selection cuts:} We require the event to have $\met>350\,$GeV and contain exactly one lepton, one b-jet with $p_T>150\,$GeV and no additional b-jet with $p_T>50\,$GeV.   To remove events with large $\met$ due to mis-measurements of jet energy, we require $|\phi_{\rm MET}-\phi_j|>0.3$ for any jet with $p_T>100\,$GeV.  We require the transverse mass of the lepton $M_T>200\,$GeV in order to remove backgrounds of which the dominate source of missing energy is from the leptonically decaying $W$ ({\it e.g.} semileptonic $t\bar{t}$).  Finally, we require the variable $\mttw$ reconstructed from the event to be at least 200\,GeV.  An event is also kept if it does not contain any additional jet for $\mttw$ to be constructed.

The variable $\mttw$, proposed in Ref~\cite{Bai:2012gs}, is constructed for dileptonic $t\bar{t}$ background with one lepton not reconstructed, and has been shown to be useful in suppressing this type of background \cite{CMS-PAS-SUS-16-002, Khachatryan:2016pup}.\footnote{Other variables have also been proposed for suppressing this background, such as $am_{\rm T2}$ \cite{Aad:2014kra} and {\it topness} \cite{Graesser:2012qy}.  As their performances are somewhat similar, for simplicity we only use $\mttw$ in this paper.}    
The calculation of $\mttw$ requires one to identify the two b-jets and to know which one is on the same side as the visible lepton.  In practice, one does not have this knowledge and would usually calculate the $\mttw$ for different possible combinations and output the minimum value from these combinations.  
Here we assume the other b-jet is among the three leading non-b-tagged jets.  We then choose the combination which minimizes $\mttw$.

\begin{figure}
\centering
\includegraphics[height=6cm]{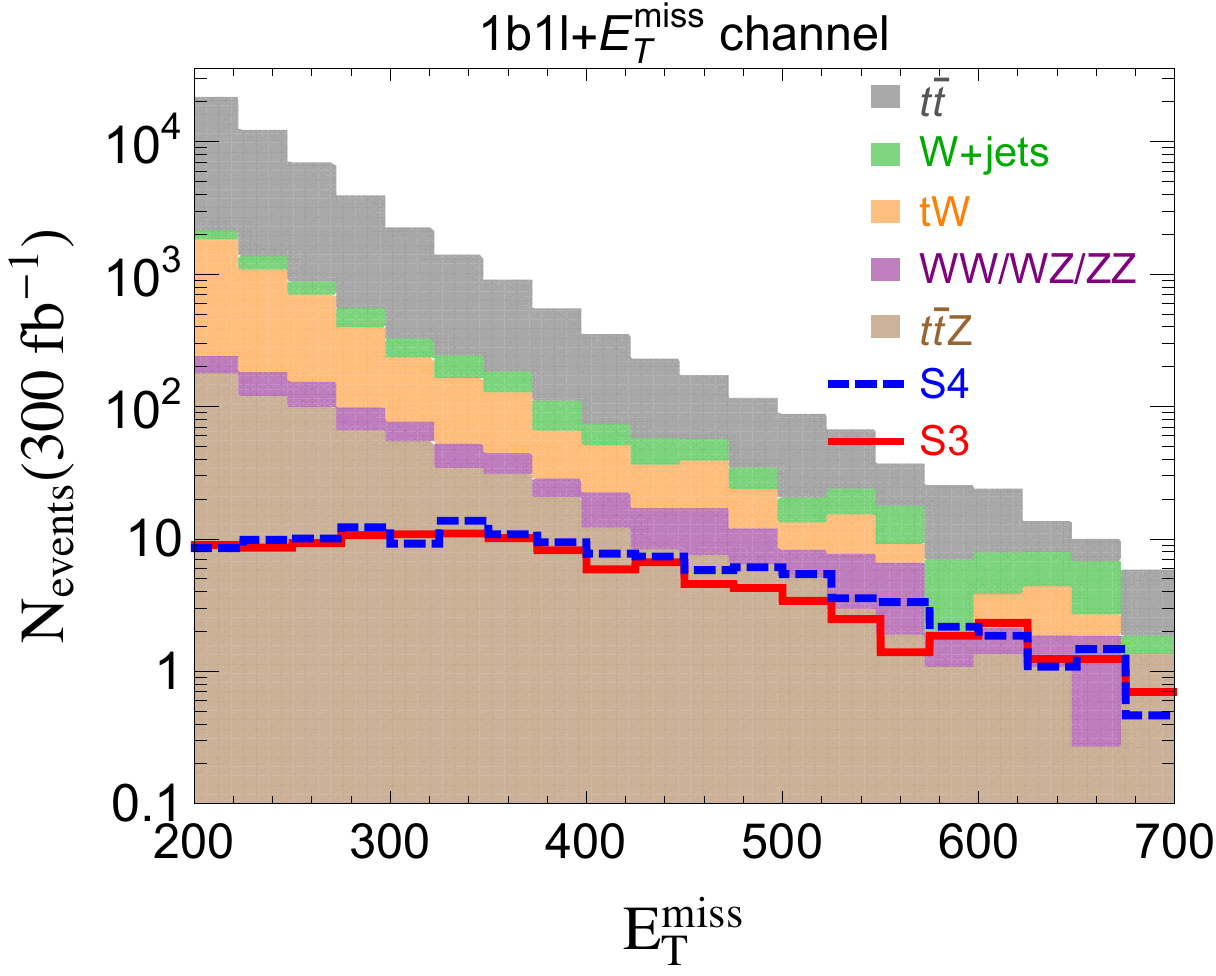} \hspace{0.5cm}
\includegraphics[height=6cm]{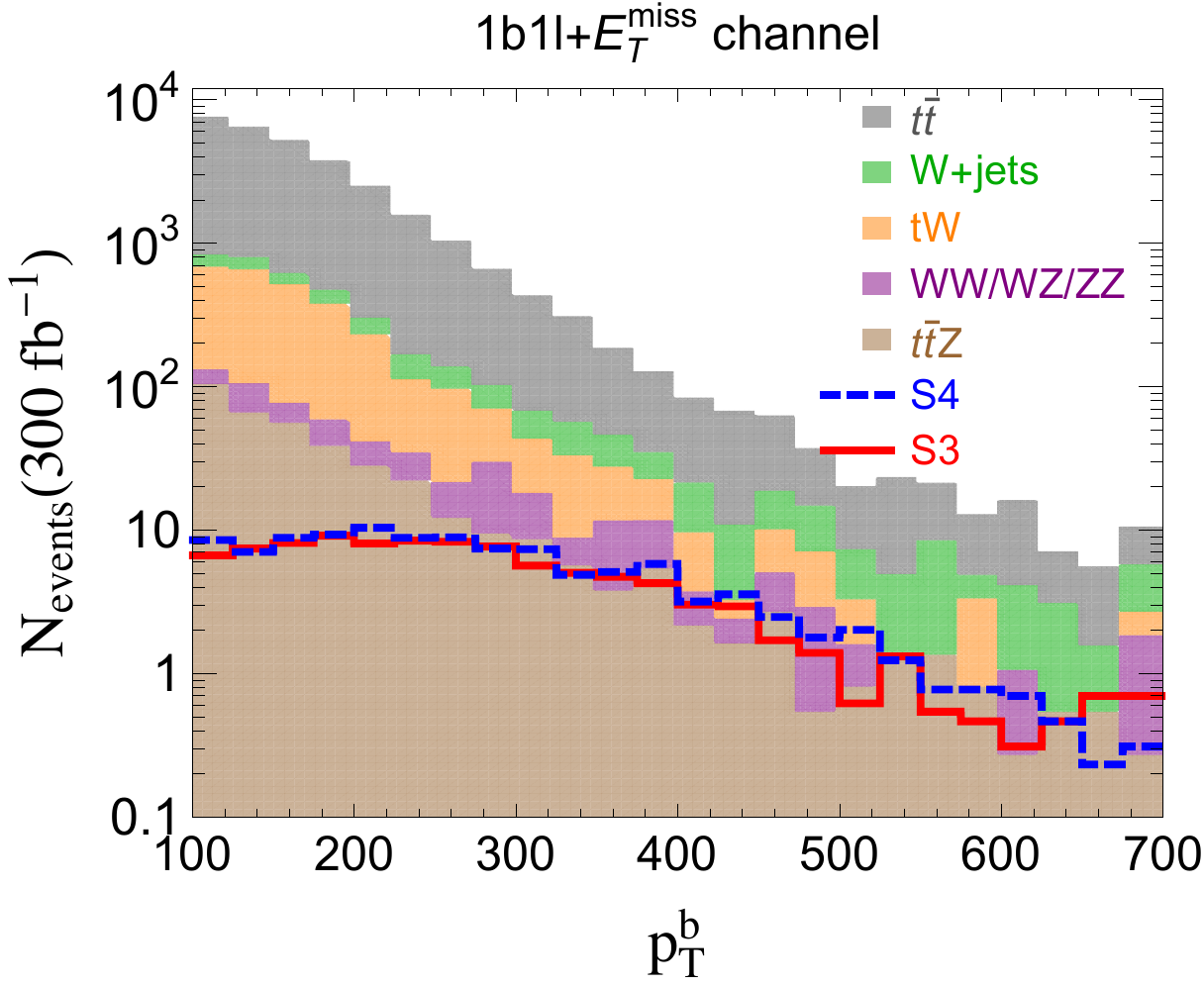}  \\ \vspace{0.5cm}
\includegraphics[height=6cm]{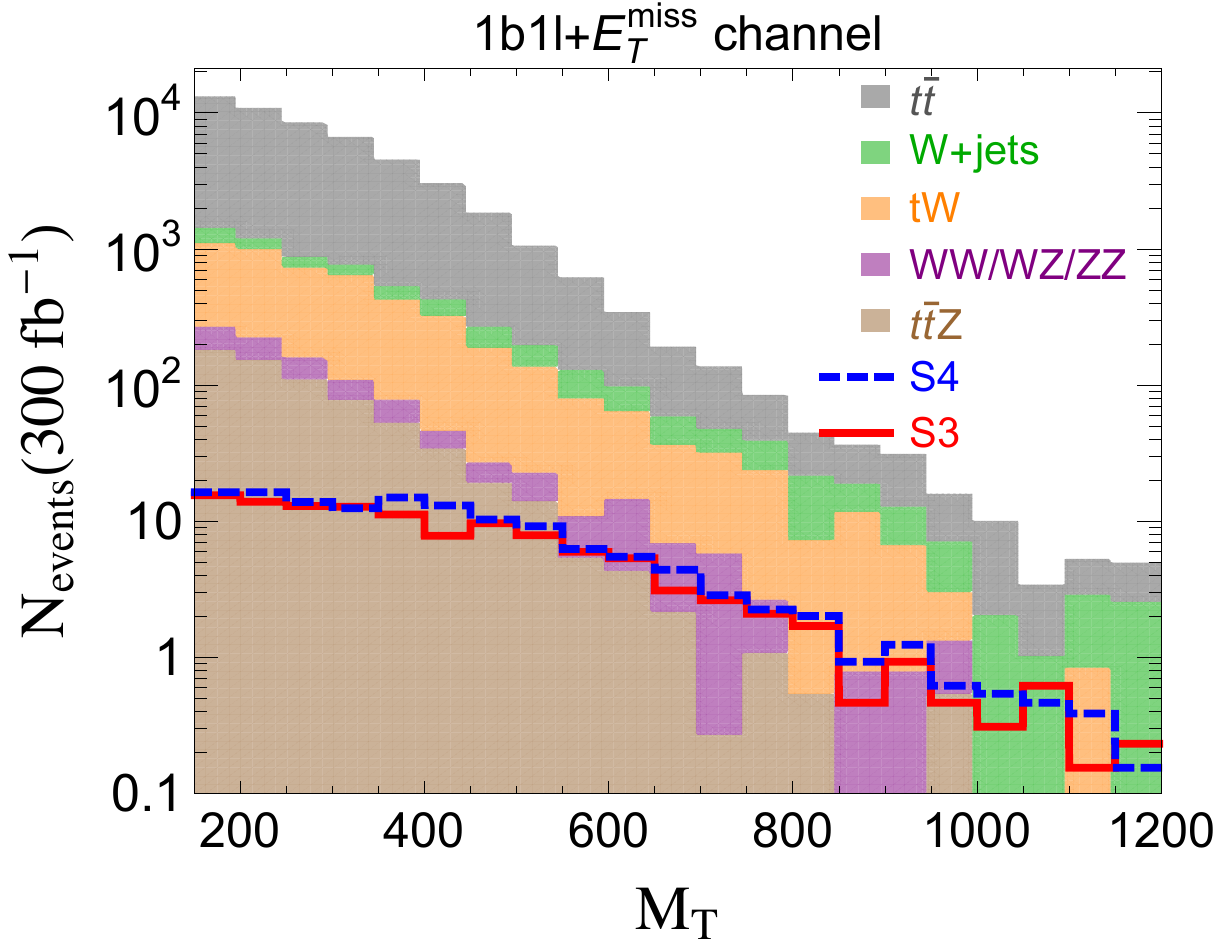} \hspace{0.5cm}
\includegraphics[height=6cm]{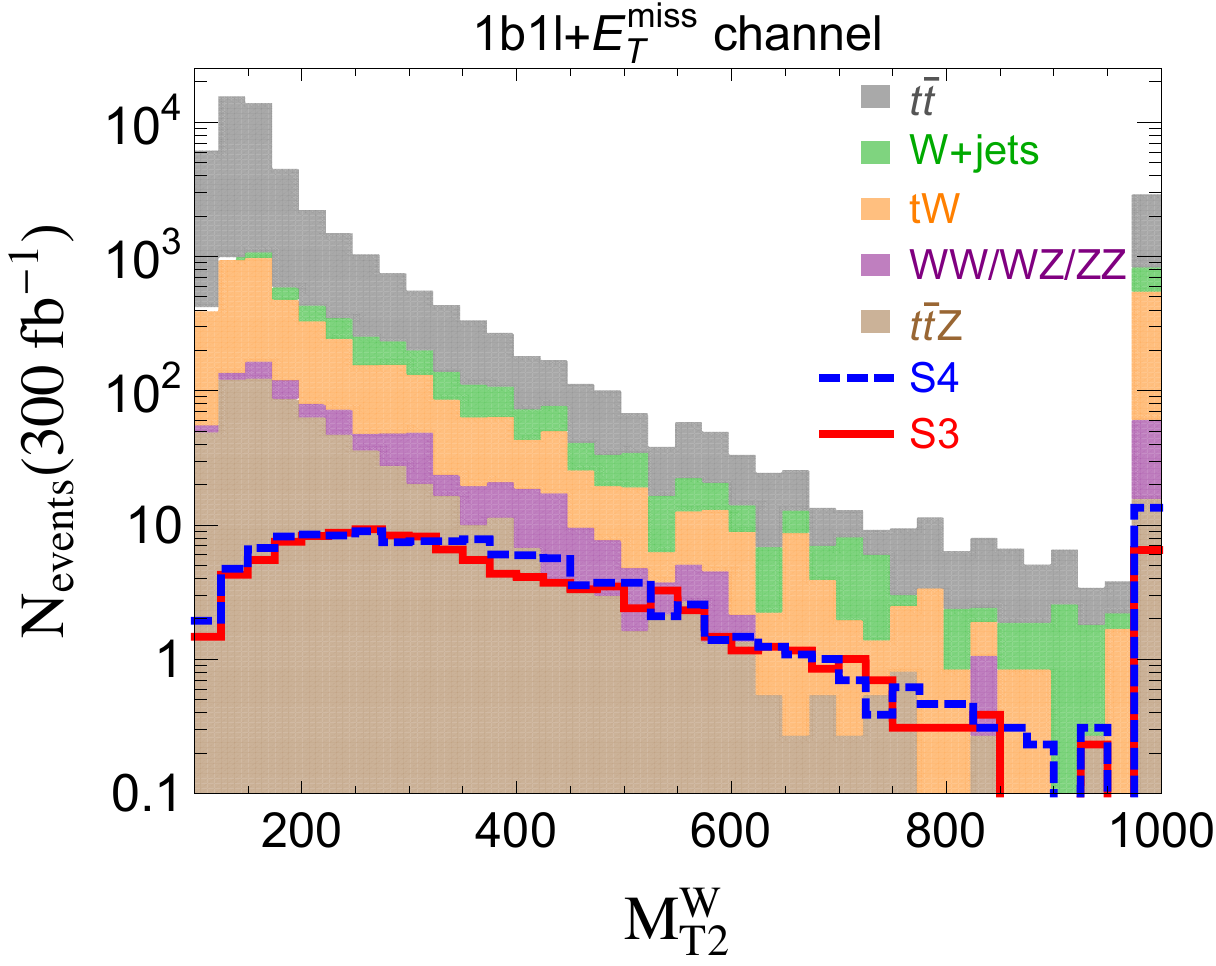}
\caption{Distributions of $\met$ ({\bf top left}), $p^b_T$ ({\bf top right}), $\mt$ ({\bf bottom left}) and $\mttwo$ ({\bf bottom right}) of the \chsemi channel for signal sample S3 and the major backgrounds.  To illustrate the usefulness of the variables, the cuts \{$\met>350\,$GeV, $p^b_T>150\,$GeV, $\mt>200\,$GeV, $\mttw>200\,$GeV\} are replaced by looser cuts \{$\met>200\,$GeV, $p^b_T>50\,$GeV, $\mt>150\,$GeV, $\mttw>0\,$GeV\}.  For the $\mttw$ distribution, events for which $\mttw$ cannot be constructed below 1\,TeV are stacked on the last bin.  The number of events correspond to $300\,{\rm fb}^{-1}$ at the 13\,TeV LHC.
}
\label{fig:semilep}
\end{figure}

The distributions of $\met$, $p^b_T$, $\mt$ and $\mttw$ are shown in Fig.~\ref{fig:semilep} for signal S1 and the major backgrounds. 
For the $\mttw$ distribution, events for which $\mttw > 1$\,TeV are stacked on the last bin.  The usefulness of $\mttw$ can be seen in Fig.~\ref{fig:semilep}, as the number of background events, in particular for $t\bar{t}$, falls sharply with $\mttw$ above the top mass.     The numbers of signals and backgrounds after the selection cuts and the corresponding $s/\sqrt{b}$ for $300\,{\rm fb}^{-1}$ data are shown in Table~\ref{tab:sig1l}.  For the \chsemi channel, the reach is also not very sensitive to the decay channel of stop.

\begin{table}
\centering
\begin{tabular}{c|cc} 
  & \# of events ($300\,{\rm fb}^{-1}$) & $s/\sqrt{b}$  \\ \hline 
S3 &  35  &  1.8   \\
S4 &  43  &  2.2  \\ \hline\hline
$t\bar{t}$ & 207  &  \\
$W+\,$jets& 84 &  \\
 $tW$  & 43   &    \\ 
$WW/WZ/ZZ$  &  28  &  \\
 $t\bar{t}Z$   &  26  &   \\ \hline
 total SM  &  389  &  
\end{tabular}
\caption{ Number of events of signal and backgrounds and the corresponding $s/\sqrt{b}$ after all of the selection cuts for the \chsemi channel with $300\,{\rm fb}^{-1}$ data.  The details of signal samples S3\,\&\,S4 are listed in Table~\ref{tab:s1234}.  All the generated backgrounds are included in the row ``total SM."      
}
\label{tab:sig1l}
\end{table}
%


\section{Reach at the 13\,TeV LHC}
\label{sec:reach}

We scan over the signal parameter space to determine the reach of the \chdi and \chsemi channels at the 13\,TeV LHC, assuming an integrated luminosity of $300\,{\rm fb}^{-1}$.  For comparison, we also include the results of the conventional search channel of the sbottom, \chbb\!\!, which has the best reach if the dominant decay of sbottom is $\tilde{b}\to b \, \chi$.  To estimate the reach of the \chbb channel, we adopt the cuts in Ref.~\cite{Aaboud:2016nwl} for signal region {\bf SRA450}, which has the best reach if the mass gap between sbottom and neutralino is large.  We have checked that the total number of backgrounds after the selection cuts, if normalized to $3.2\, {\rm fb}^{-1}$, is in good agreement with Ref.~\cite{Aaboud:2016nwl}.  We use the asymptotic formula for the significance in Ref.~\cite{Cowan:2010js} (also adopted by Ref.~\cite{Cheng:2016mcw, Cheng:2016npb}),
\begin{equation}
\sigma = \sqrt{2 \left[ (s+b)\log{(1+s/b)} -s  \right ] }  \,,
\end{equation}
which reduces to the usual $s/\sqrt{b}$ in the limit $b\gg s$.  
While the optimal values of the selection cuts depend on the signal spectrum, for simplicity we fix the cuts as in Section~\ref{sec:study}.  In particular, for the \chdi channel the cuts we choose are relatively conservative to maintain a sufficiently large simulated signal sample.  A more sophisticated optimization method could further improve the reach of the searches.

\begin{figure}[t]
\centering
\includegraphics[height=8cm]{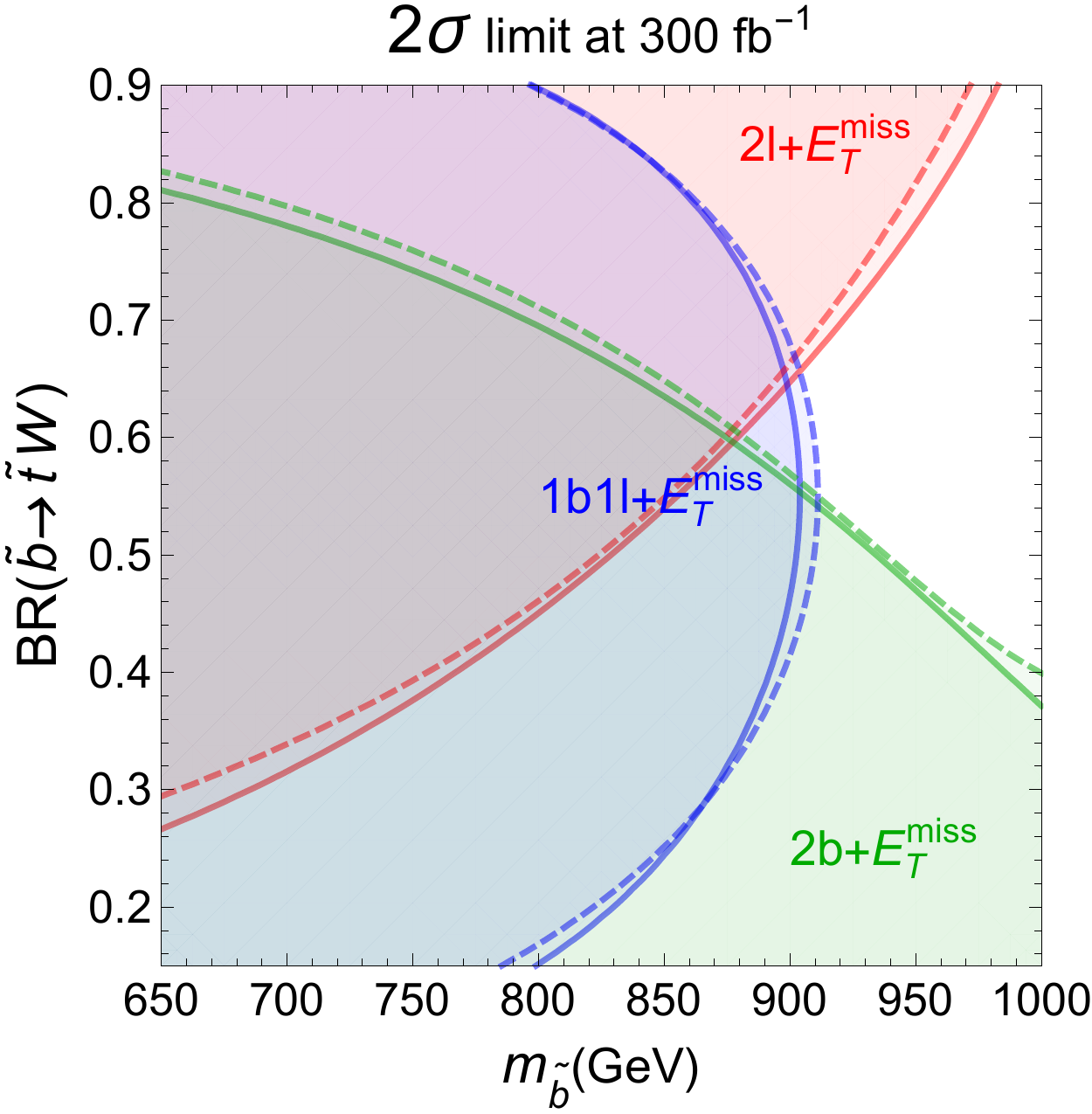} 
\includegraphics[height=8cm]{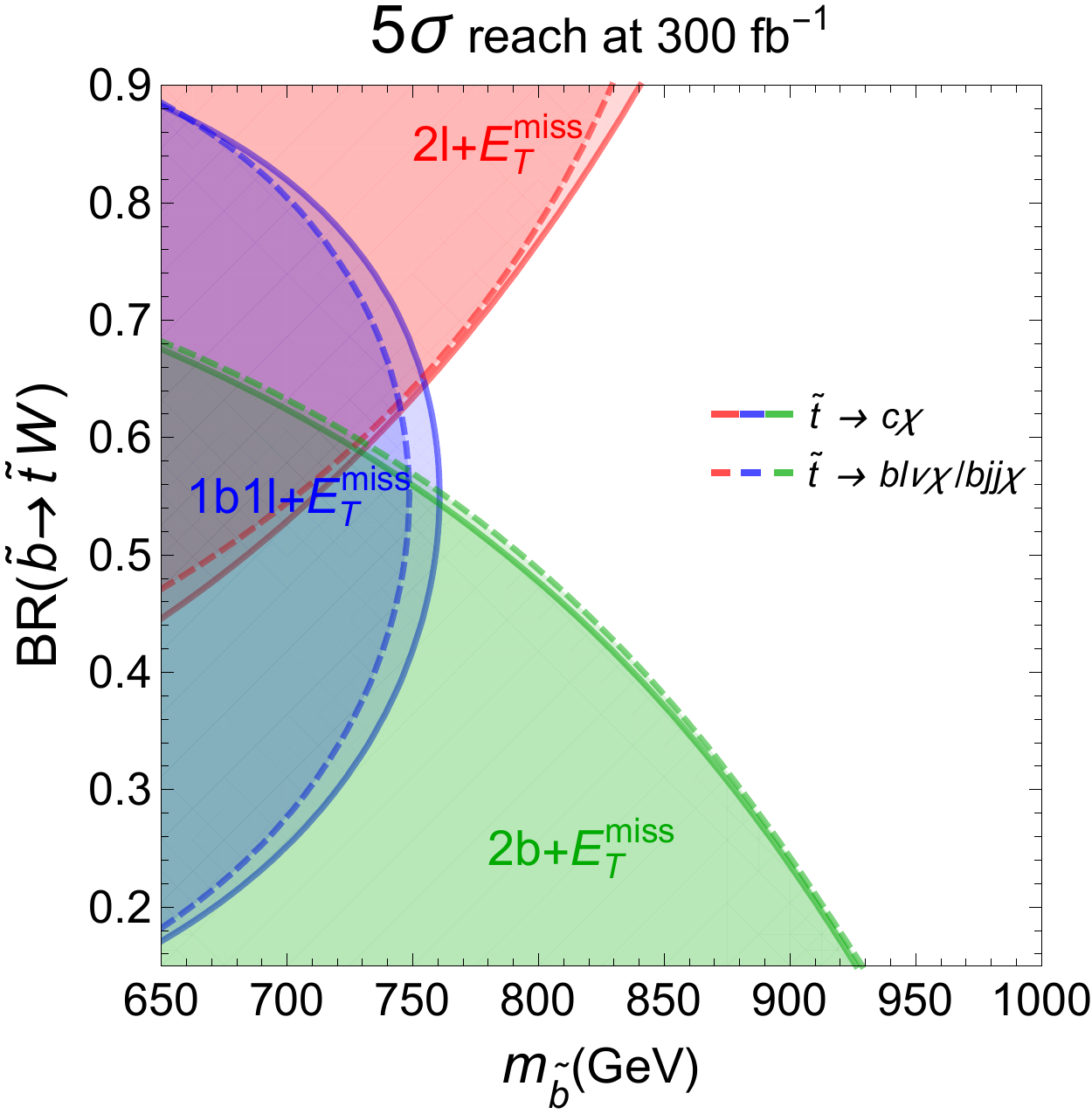}
\caption{Expected $2\,\sigma$ limits (left) and $5\,\sigma$ reaches (right) in the $(m_{\tilde{b}},~{\rm BR}(\tilde{b}\to\tilde{t}\,W))$ plane from the 13\,TeV LHC with $300\,{\rm fb}^{-1}$ data, assuming $m_{\tilde{b}}-m_{\tilde{t}}=400$\,GeV and $m_{\tilde{t}}-m_{\chi}=30$\,GeV.  The red, blue and green contours indicate the regions excluded by the \chdi\!\!, \chsemi and \chbb channels, respectively.  The solid (dashed) curves corresponds to the stop decay $\tilde{t}\to c\,\chi$ ($\tilde{t}\to bW^*\chi \to bl\nu \chi / bjj \chi$) with $100\%$\,BR.  
The contours are obtained by scanning the $(m_{\tilde{b}},~{\rm BR}(\tilde{b}\to\tilde{t}\,W))$ plane with a grid spacing of (20\,GeV, 0.05) and fitting the points to a 2-dimensional polynomial.
}
\label{fig:brmsb}
\end{figure}

In Fig.~\ref{fig:brmsb}, we show the expected exclusion regions for the three channels in the $(m_{\tilde{b}},~{\rm BR}(\tilde{b}\to\tilde{t}\,W))$ plane, assuming $m_{\tilde{b}}-m_{\tilde{t}}=400$\,GeV and $m_{\tilde{t}}-m_{\chi}=30$\,GeV.  On the left panel, the red, blue and green contours indicate the 2-sigma limits of the \chdi, \chsemi and \chbb channels, respectively, and the corresponding shaded regions are excluded at 95\% confidence level (CL).  On the right panel, the 5-sigma reaches are shown instead.  The contours are obtained by scanning the $(m_{\tilde{b}},~{\rm BR}(\tilde{b}\to\tilde{t}\,W))$ plane with a grid spacing of (20\,GeV, 0.05). A fit to a 2-dimensional polynomial was performed to reduce the unphysical fluctuations due to the statistical uncertainties of the simulations.  We also checked manually that the fitted curves are in good agreement with the grid of data points.  For the solid curves, we assume the stop only decays to charm and neutralino, $\tilde{t}\to c\,\chi$; for the dashed curves, we assume that the stop only goes through the 4-body decay, $\tilde{t}\to bW^*\chi \to bl\nu \chi / bjj \chi$.  As we expected, the stop decay channel has a small impact on the reach.  The complementarity of different channels is well demonstrated in Fig.~\ref{fig:brmsb}.  The \chdi (\chbb) channel has the best reach if the decay $\tilde{b}\to\tilde{t}\,W$ ($\tilde{b}\to b \, \chi$)  is dominant, and the \chsemi channel has a better reach if the branching ratio of the two decay channels are comparable.  We also found that the \chbb channel has rather good reaches, comparable to the reach of the \chsemi channel even for ${\rm BR}(\tilde{b}\to\tilde{t}\,W)\sim0.5$.  Nevertheless, the \chsemi channel could still significantly improve the overall significance (of all channels combined) and impose constraints on the stop mass.

\begin{figure}
\centering
\includegraphics[height=7.5cm]{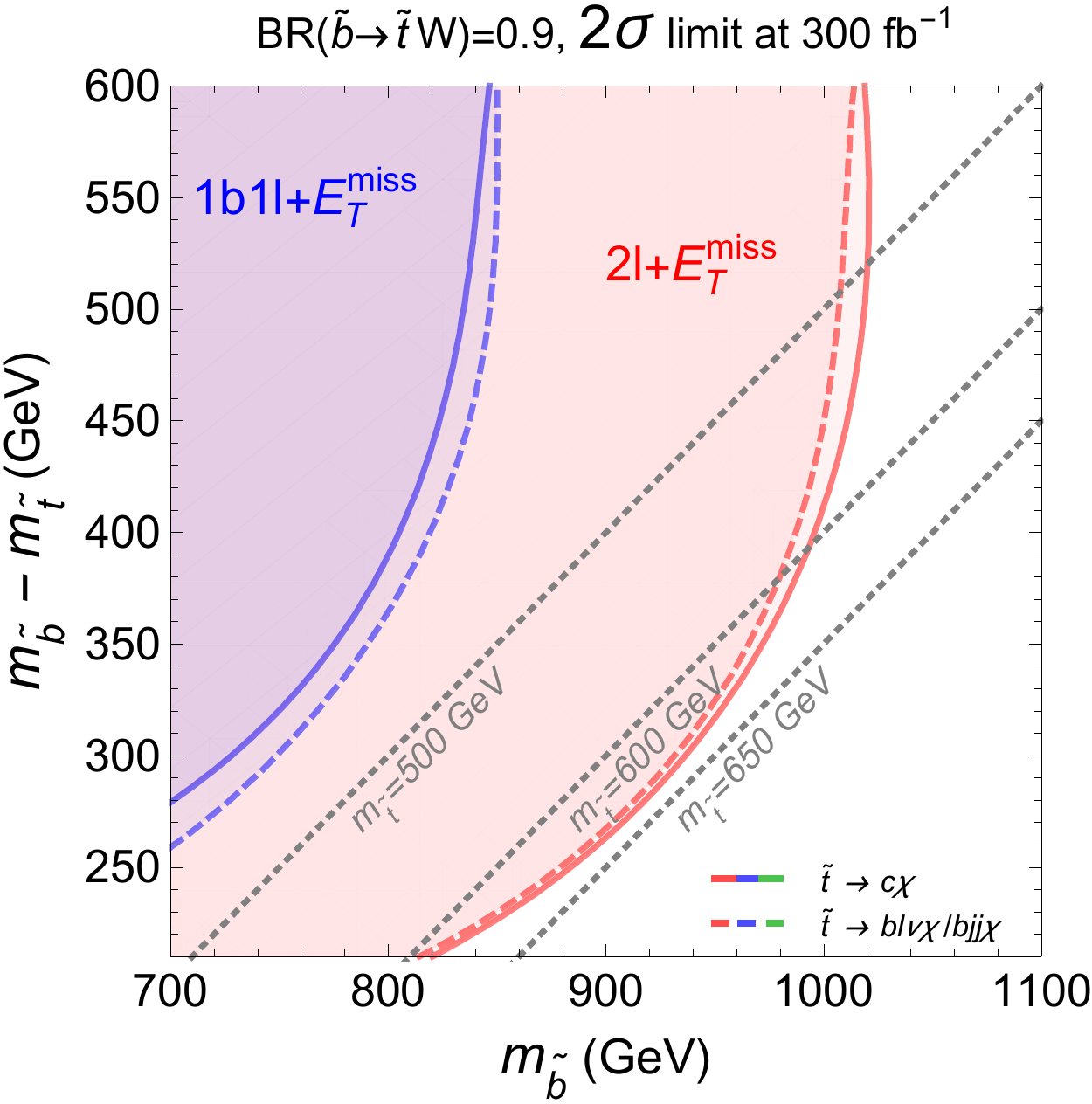} 
\includegraphics[height=7.5cm]{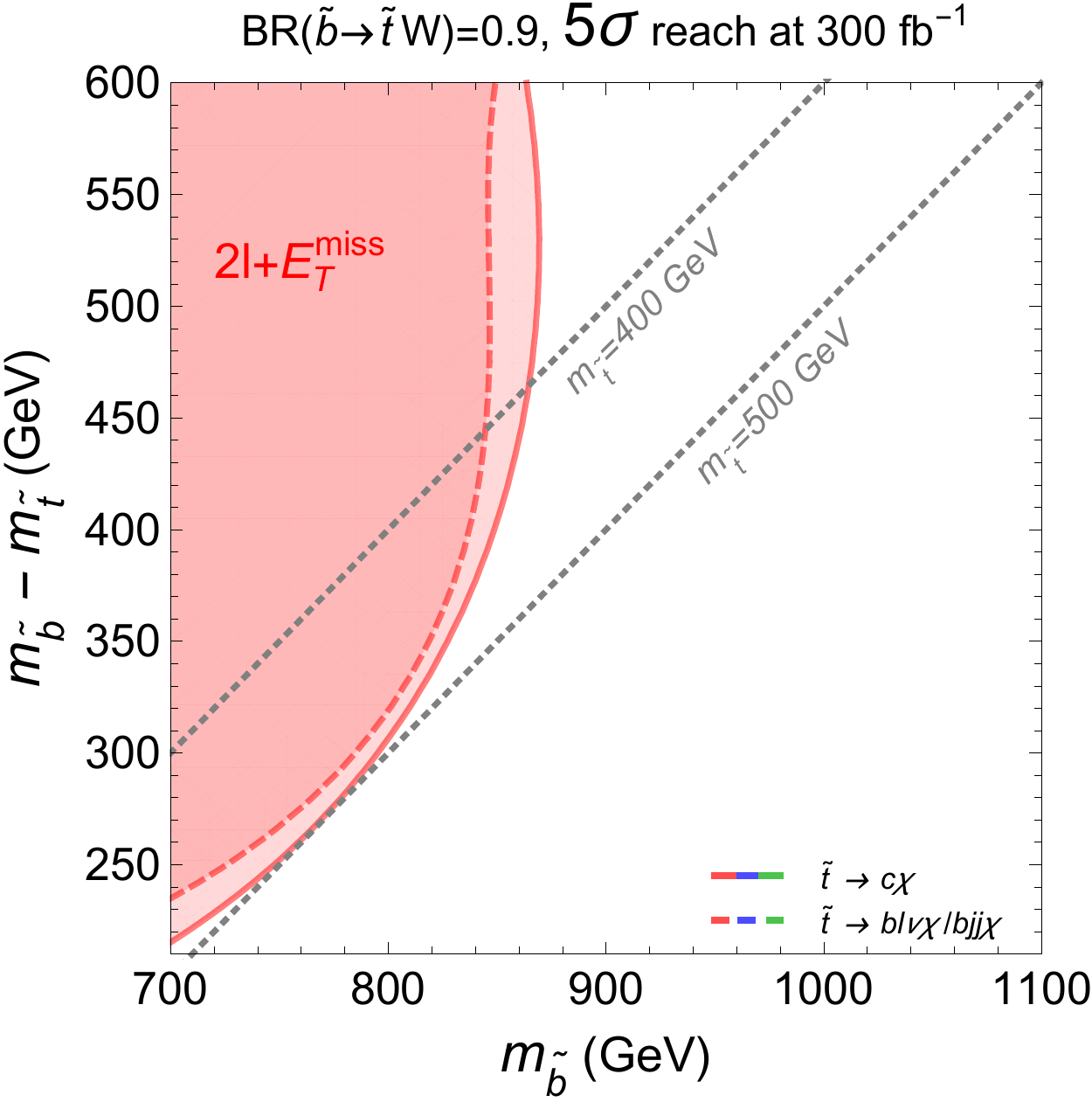} \\ \vspace{0.7cm}
\includegraphics[height=7.5cm]{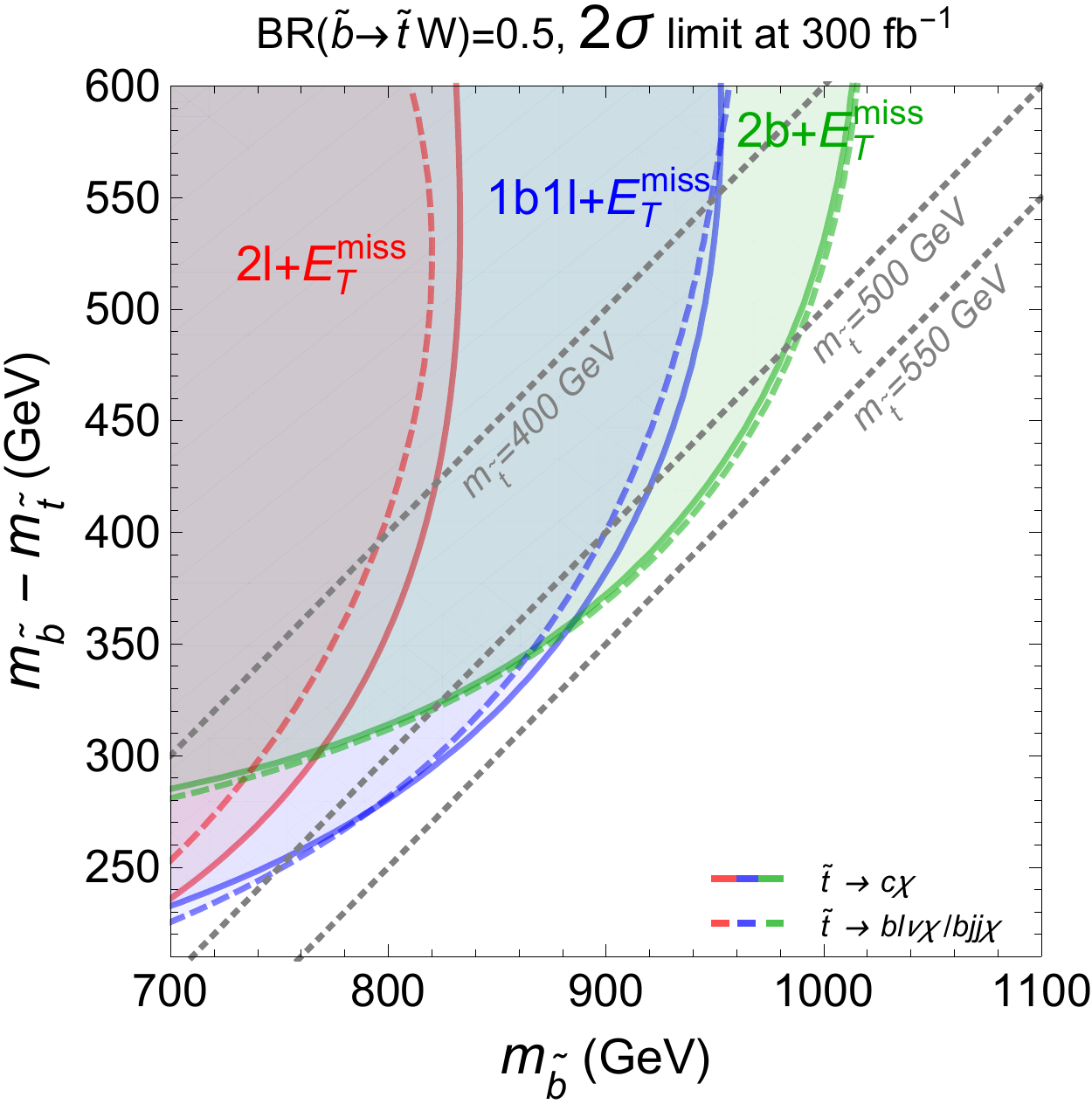}
\includegraphics[height=7.5cm]{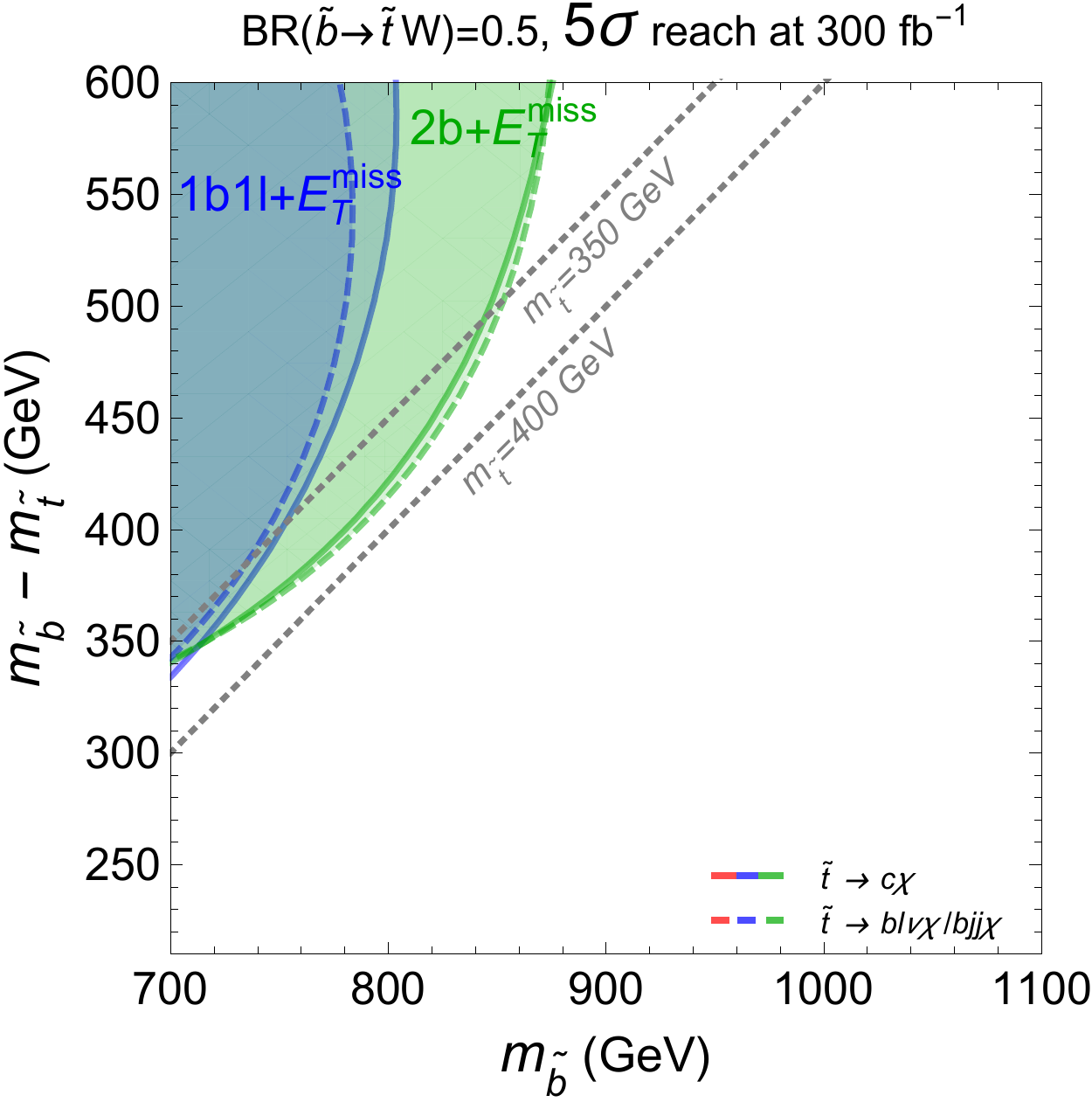}
\caption{Expected $2\,\sigma$ limits (left) and $5\,\sigma$ reaches (right) in the $(m_{\tilde{b}},~m_{\tilde{b}}-m_{\tilde{t}})$ plane from the 13\,TeV LHC with $300\,{\rm fb}^{-1}$ data, assuming $m_{\tilde{t}}-m_{\chi}=30$\,GeV.  The top (bottom) panel assumes ${\rm BR}(\tilde{b}\to\tilde{t}\,W) = 0.9~(0.5)$.  The red, blue and green contours indicate the regions excluded by the \chdi\!\!, \chsemi and \chbb channels, respectively. The solid (dashed) curves corresponds to the stop decay $\tilde{t}\to c\,\chi$ ($\tilde{t}\to bW^*\chi \to bl\nu \chi / bjj \chi$) with $100\%$\,BR.  The dotted diagonal lines correspond to constant values of $m_{\tilde{t}}$.  
The contours are obtained by scanning the $(\msb,~\msb-\mst)$ plane with a grid spacing of (20\,GeV,~30\,GeV) and fitting the points to a 2-dimensional polynomial.
}
\label{fig:mm}
\end{figure}

To determine the bounds on masses of sbottom and stop, we also show the 2-sigma limits and 5-sigma reaches in the $(\msb,~\msb-\mst)$ plane in Fig.~\ref{fig:mm} for ${\rm BR}(\tilde{b}\to\tilde{t}\,W) = 0.9$ (top panel) and $0.5$ (bottom panel), assuming $m_{\tilde{t}}-m_{\chi}=30$\,GeV.  Similar to Fig.~\ref{fig:brmsb}, the contours are obtained by scanning the $(\msb,~\msb-\mst)$ plane with a grid spacing of (20\,GeV,~30\,GeV) and fitting the points to a 2-dimensional polynomial.  A few benchmarks of stop masses are also shown, which correspond to diagonal lines in the $(\msb,~\msb-\mst)$ plane.  For ${\rm BR}(\tilde{b}\to\tilde{t}\,W) = 0.9$, it is clear that the \chdi channel has the best reach, and stop masses up to $\sim 600\,$GeV can be excluded for $\msb \lesssim 1\,$TeV.    
For ${\rm BR}(\tilde{b}\to\tilde{t}\,W) = 0.5$, the \chsemi and \chbb channels have comparable reaches, with the \chsemi (\chbb\!\!) channel having better constraints on $\mst$ in the regions with smaller (larger) $\msb$.  However, it should be noted that the \chbb channel does not direct constrain $m_{\tilde{t}}$, and the exclusion region shown in Fig.~\ref{fig:mm} is based on the assumption $m_{\tilde{t}}-m_{\chi}=30$\,GeV.  For the \chsemi channel, stop masses up to $\sim 500\,$GeV can be excluded for $\msb \lesssim 900\,$GeV.  It should also be noted that in obtaining the constraints we have assumed a sufficient mass gap between the sbottom and the stop.  If the mass gap is small, the search strategy can be drastically different, in particular in the region $\msb - \mst \lesssim m_W$.  Further studies are required to determine the collider reach in this region.

Comparing to the reach of the 
direct stop search, the recent results from the ATLAS mono-jet search has excluded stop masses below 323\,GeV with $3.2\,{\rm fb}^{-1}$ data at $\sqrt{s}=13\,$TeV, assuming $m_{\tilde{t}} - m_\chi \approx 5\,$GeV  \cite{Aaboud:2016tnv}.  This already surpasses the constraints from the 8\,TeV run \cite{Aad:2014nra, Khachatryan:2016pxa}.  
CMS conducted a search  with $2.3\,{\rm fb}^{-1}$ data at $\sqrt{s}=13\,$TeV using the $\alpha_{\rm T}$ variable which can exclude stop masses up to 400\,GeV assuming $m_{\tilde{t}} - m_\chi \approx 10\,$GeV \cite{Khachatryan:2016dvc}.  In both searches, the bounds on stop mass are also significantly weaker for slightly larger values of $m_{\tilde{t}} - m_\chi$.
In Ref.~\cite{Low:2014cba}, it is estimated that the high luminosity LHC with $3000\,{\rm fb}^{-1}$ data at $\sqrt{s}=14\,$TeV is required for the bounds on stop mass from mono-jet search to reach $\sim 500$\,GeV, assuming the stop is in the coannihilation region.  While the constraints from mono-jet searches do not rely on the properties of sbottom and are hence more robust, the search with sbottom decays could potentially have a much better reach.  The two searches are also complementary; if a significant excess is found in the \chdi or \chsemi channel, one may also expect a mild excess in the mono-jet search if the excess comes from a light stop in the coannihiliation region.


\section{Conclusion}
\label{sec:conclusion}

A light stop with mass almost degenerate with the lightest neutralino is an appealing SUSY scenario.  It could evade the bounds of traditional stop searches and hence reduce the tension between naturalness and the current LHC results, while also having interesting implications for bino dark matter.     
In this paper, we propose a novel way of probing such stop particles by searching for it from sbottom decays, under the assumptions that the sbottom is not too heavy and has a significant branching ratio of decaying into a stop and a $W$~boson ($\tilde{b} \to \tilde{t}\,W$).  Such assumptions are favored by naturalness and Higgs mass considerations.  In this scenario, the constraints on the masses of stop and sbottom from the traditional searches are weak.  
We show that a dedicated search for a sbottom pair with one or both sbottom decaying to stop and $W$ at the 13\,TeV LHC could impose strong constraint on this scenario, hence making it the optimal search channel.  
Assuming $\mst - m_\chi \approx 30\,$GeV, if the decay $\tilde{b}\to\tilde{t}\,W$ is dominant, the channel with final states \chdi has the best reach, and can exclude stop masses up to $\sim600\,$GeV with $300\,{\rm fb}^{-1}$ data if the sbottom is below 1\,TeV;  
if the sbottom decays to either $\tilde{t}\,W$ or $b\,\chi$ with comparable branching ratios, the channel with final states \chsemi  has a better reach and could exclude the stop with mass up to $\sim500\,$GeV with $300\,{\rm fb}^{-1}$ data if the sbottom is below 900\,GeV.  While the results rely on the properties of the sbottom, the reaches are potentially much better than the one from direct searches of stop with mono-jet + $\met$ final states, which could only reach up to $\sim500\,$GeV with $3000\,{\rm fb}^{-1}$ data at $\sqrt{s}=14\,$TeV.  The traditional search channel of sbottom with final states {\boldmath $2b+\met$} is also complementary to the ones we propose.  Together, these searches can cover a wide range of model parameter space and provide valuable information on the status of SUSY.

There are other interesting scenarios not explored in this paper but may worth further investigation.  It is possible that the chargino or second neutralino are lighter than the sbottom, making its decay more complicated \cite{Han:2015tua}.  In this case, searching for the asymmetric decay chains with one sbottom decaying to $\tilde{t}\,W$, the other decaying to $t\chi^\pm$ or $b\chi_2$ could be useful.  For larger values of the stop-neutralino mass gap, the stop decay products become more visible and it might be useful to look at channels with multiple b-jets or multiple leptons \cite{Cheng:2016npb}, or try to tag the charm quark from stop decay \cite{Choudhury:2012kn, Belanger:2013oka, Aad:2014nra}.  On the other hand, if the mass gap is smaller, the stop decay could exhibit displaced vertex, which can help reduce SM background in both the mono-jet search and the search with sbottom decays.  It is also complementary to search for the lighter stop from the decays of the heavier stop \cite{Perelstein:2007nx, Ghosh:2013qga, Khachatryan:2014doa}.  

Our study serves as a proof of concept.  A search carried out by the LHC experimental groups is desired to fully determine the reach of the  proposed channels.  Since the \chdi channel has been used to search for sleptons and electroweakinos, and the conventional search of stop in the semileptonic channel is very similar to the \chsemi channel we studied, reinterpretation of those search results can already lead to interesting reach.  At the same time, optimizing the searches with these new channels in mind is needed to realize their full potential.  While the current data is still not very constraining, in the future it is straightforward to interpret the results of conventional searches in these two channels in terms of constraints on the scenario studied in this paper.  If deviations from the SM is observed, it is non-trivial to discriminate different new physics scenarios that leads to similar signals, and the comparisons between different search channels are important.


\subsection*{Acknowledgments}
We would like to thank Zhen Liu for useful discussions. HA is supported by the Walter Burke Institute at Caltech and by DOE Grant de-sc0011632.  JG is supported by the International Postdoctoral Exchange Fellowship Program between the Office of the National Administrative Committee of Postdoctoral Researchers of China (ONACPR) and DESY.  JG would also like to express a special thanks to the Mainz Institute for Theoretical Physics (MITP) for its hospitality and support.  LTW is supported by DOE grant DE-SC0013642.


\providecommand{\href}[2]{#2}\begingroup\raggedright\endgroup


\end{document}